\documentclass[prb,twocolumn,showpacs,amsmath,amssymb]{revtex4}
\usepackage{graphicx,ulem}
\usepackage{enumerate}
\newcommand \beq{\begin{eqnarray}}
\newcommand \eeq{\end{eqnarray}}

\newcommand{\bfr}{\mathbf{r}}
\newcommand{\bfk}{\mathbf{k}}
\newcommand{\bfK}{\mathbf{K}}
\newcommand{\bfs}{\mathbf{s}}

\newcommand{\ahc}{a_{_\mathrm{Hc}}}
\newcommand{\atau}{a_{\tau'}}
\newcommand{\vf}{v_{_F}}
\newcommand{\feff}[1]{F_{\mathrm{eff}}^{( #1 )}}

\begin{document}

\title{Chiral symmetry restoration in monolayer graphene 
induced by Kekul\'e distortion}
\author{Yasufumi Araki}
\affiliation{
Department of Physics, The University of Tokyo, Tokyo 113-0033, Japan
}

\begin{abstract}
We propose a chiral symmetry restoration mechanism
in monolayer graphene,
in analogy with the strongly coupled gauge theory.
The chiral (sublattice) symmetry of graphene,
which is spontaneously broken under the effectively strong Coulomb interaction,
is restored by introducing the Kekul\'e-patterned lattice distortion externally.
Such a phase transition is investigated analytically using the techniques of strong coupling expansion
on the lattice gauge theory model, by preserving the original honeycomb lattice structure.
We discuss the relation between the chiral phase transition and the spectral gap amplitude,
and we show the modification of the dispersion relation of the quasiparticles.
\end{abstract}
\pacs{73.22.Pr,71.35.-y,11.15.Ha,11.15.Me}
\maketitle

Ever since its first experimental isolation in 2004 \cite{Novoselov_2004},
graphene has attracted a great deal of interest in the fields of particle physics
as well as materials science \cite{castro_neto_2009}.
One of its important features is that
the electrons/holes on monolayer graphene can be effectively described
as massless Dirac fermions around half filling \cite{wallace_1947,Semenoff_1984}.
Since the fermions on the (2+1)-dimensional sheet
interact with the electromagnetic field in the (3+1)-dimensional space,
one can apply
``reduced quantum electrodynamics (QED)'' \cite{Gorbar_2001}
as an effective field theory for this real material.

In reduced QED,
the chiral symmetry is suggested to be spontaneously broken
if the Coulomb interaction between fermions is effectively strong.
In graphene,
such a strong Coulomb interaction is supposed to be achieved
if the electromagnetic field is not screened by dielectric substrates,
i.e. the layer is suspended in the vacuum.
Breaking of the chiral symmetry corresponds to
the spontaneous breaking of the inversion symmetry
between two triangular sublattices,
leading to the generation of a mass gap \cite{Physics_2009}.
This mechanism is analogous to the dynamical mass generation mechanism of quarks
in quantum chromodynamics (QCD) \cite{Hatsuda_Kunihiro}.
The techniques of lattice gauge theory
have been employed, on the hypothetical square lattice,
to investigate the chiral symmetry breaking in graphene
either numerically \cite{Drut_2009} or analytically \cite{Araki_2010}.


In this paper,
we propose a mechanism that restores the ``chiral symmetry''
even under the strong Coulomb interaction.
Such a phase transition is induced by the external Kekul\'e distortion,
which is described by the alternating bond strength like the benzene molecule \cite{Viet_1994}.
The Kekul\'e distortion yields the asymmetry between two Dirac points
and gives a finite gap in the spectrum,
but does not break the chiral (sublattice) symmetry explicitly.
It can be introduced externally
by the effect of the substrate or the addition of atoms on the layer
\cite{Cheianov_2009,Farjam_2009}.
The interplay between
the breaking of the chiral symmetry and the Kekul\'e distortion
at long wavelength, i.e. in the mean field,
has not yet been understood in previous works \cite{Dillenschneider_2008}.

In order to incorporate the effect of the Kekul\'e distortion,
the effective gauge theory model with the original honeycomb lattice structure
would be helpful \cite{Chakrabarti_2009,Giuliani_2010}.
First we construct a U(1) lattice gauge theory model for graphene,
with both the spatial and temporal directions regularized on the hexagonal-prismatic lattice.
Then, we apply to this model
the strong coupling expansion techniques of the lattice gauge theory,
which has been employed widely in analyzing the properties
of strong coupling gauge theories such as QCD
\cite{Kawamoto_1981,Drouffe_1983}.
We demonstrate the chiral symmetry restoration transition
by adding the Kekul\'e distortion term to the model action.
The interplay between two orders results in the modification of the gap size,
depending on the distortion amplitude.

\begin{figure}[tb]
\begin{center}
\includegraphics[width=5.5cm]{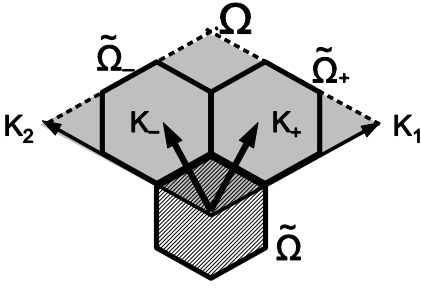}
\end{center}
\vspace{-0.5cm}
\caption{A schematic picture of the Brillouin zone corresponding to the honeycomb lattice.
The Brillouin zone $\Omega$ (gray rhombic region) is spanned by the reciprocal lattice vectors $\bfK_1$ and $\bfK_2$.
If the Kekul\'{e} distortion pattern [Eq.(\ref{eq:k-pattern})]
 is introduced,
$\Omega$ is split into three hexagonal cells:
$\tilde{\Omega}$ and $\tilde{\Omega}_\pm$,
surrounding $\bfk=0$ and $\bfK_\pm$ (Dirac points) respectively.
}
\label{fig:brillouinzone2}
\end{figure}

Noninteracting electrons on the monolayer graphene with the perfect honeycomb lattice symmetry
 are described by
the conventional tight-binding Hamiltonian
\begin{equation}
H = h \sum_{\mathbf{r} \in A} \sum_{i=1,2,3} \left[ a^\dag(\bfr) b(\bfr+\bfs_i) + b^\dag(\bfr+\bfs_i) a(\bfr) \right],  \label{eq:tight-binding}
\end{equation}
with the nearest-neighbor hopping amplitude $h$.
The operators $a^\dag \, (a)$ and $b^\dag \, (b)$ create (annihilate) a fermion
on the triangular sublattices A and B, respectively.
$\bfs_i \, (i=1,2,3)$ is the orientation vector pointing from an A-site to a neighboring B-site,
with the lattice spacing $|\bfs_i| \equiv \ahc = 1.42$\AA.
This Hamiltonian yields the energy eigenvalue $E(\bfk) = \pm h|\Phi(\bfk)|$,
where $\Phi(\bfk) \equiv \sum_{i=1,2,3} e^{i\bfk\cdot\bfs_i}$.
$E(\bfk)$ vanishes at two Dirac points $\bfK_\pm$ in the Brillouin zone $\Omega$ (see Fig.\ref{fig:brillouinzone2}).
By neglecting quadratic and higher order terms in momentum,
$\Phi$ can be written around the Dirac points as $\Phi(\bfK_\pm + \bfk) = (3/2)\ahc (\pm k_x +i k_y) +O(k^2)$,
which we call here the ``linear approximation''.
Thus, the dispersion relation can be linearized around the Dirac points,
 with the Fermi velocity
$\vf= (3/2)\ahc h$ about 300 times smaller than the speed of light \cite{wallace_1947}.
An effective field theory description has been established in continuum,
with a 4-component Dirac fermion 
$\psi = [a(\bfK_+ +\bfk),a(\bfK_- +\bfk),b(\bfK_+ +\bfk),b(\bfK_- +\bfk)]^T$ \cite{Semenoff_1984}.

From the tight-binding Hamiltonian in Eq.(\ref{eq:tight-binding}),
we can construct a Euclidean action for fermions
with the original honeycomb lattice structure.
As done in Ref.\onlinecite{Araki_2010},
we perform a temporal scale transformation $\tau \rightarrow \tau'/\vf$,
so that the effective Fermi velocity in the action can be rendered to be unity.
Here we discretize the temporal direction with the lattice spacing $\atau$,
which is set to be comparable to the spatial lattice spacing $\ahc$.
(In this paper, we set $\atau$ equal to $\ahc$ for convenience of calculation.)
A pair of fermion doublers are inevitably generated in the temporal direction \cite{Nielsen_Ninomiya_1981},
which we reinterpret as the spin (up/down) degrees of freedom.

In order to incorporate the interaction between the fermions and the electromagnetic field,
we require the local U(1) gauge invariance, with the compact link variables
$U_{\tau'}(\bfr,\tau') = \exp\left[i e \int_{\tau'}^{\tau'+\atau} d\tau' A_4\right]$
and $U_i(\bfr,\tau') = \exp\left[i e \int_{\bfr}^{\bfr+\bfs_i} d\bfr \cdot \mathbf{A}\right]$
($e$ is the electric charge);
thus we obtain a ``gauged'' honeycomb lattice model
\begin{eqnarray}
S_F &=& \!\!\! \frac{1}{2} \!\! \sum_{\bfr \in A;\tau'} \!\!\! \left[a^\dag(x)U_{\tau'}(x)a(x+\atau)-\mathrm{H.c.}\right] \label{eq:lattice-action} \\
 && \!\!\!\!\!\! +\frac{1}{2} \!\!\! \sum_{\bfr \in B;\tau'} \!\!\! \left[b^\dag(x)U_{\tau'}(x)b(x+\atau)-\mathrm{H.c.}\right] \nonumber \\
 && \!\!\!\!\!\!\!\!\!\!\!\! +\frac{\atau h}{\vf} \sum_{\mathbf{r} \in A,\tau'} \sum_{i=1}^{3} \left[ a^\dag(x) U_i(x) b(x+\bfs_i) + \mathrm{H.c.} \right], \nonumber
\end{eqnarray}
where $x$ denotes the (2+1)-dimensional position $(\bfr,\tau')$.
Here we take $a^\dag$ and $b^\dag$ ($a$ and $b$) as the Grassmann fields
corresponding to the creation (annihilation) operators
of the fermionic quasiparticles around half-filling.
Kinetic term of the gauge field, $S_G$,
can also be discretized on the honeycomb lattice with the U(1) link variables.
$S_G$ becomes proportional to the inverse of the effective Coulomb coupling strength,
$\beta \equiv \epsilon \vf / e^2 = 4\pi\epsilon/\alpha_{_\mathrm{QED}}$,
where $\epsilon$ is the dielectric constant of the surrounding material
and $\alpha_{_\mathrm{QED}} \simeq 1/137$ is the fine-structure constant of QED.

This lattice model reproduces {the ``reduced QED''-like model} \cite{Gorbar_2002,Son_2007}
in the continuum limit.
The fermions (quasiparticles) propagate much slower than the electromagnetic field (photons) (i.e. $\vf \ll c$),
so that the retardation (magnetic) effect of the electric field
can safely be neglected.
Therefore, the spatial link variables $U_i$ can be set to unity here,
which we call the ``instantaneous approximation''.

Now we can perform the strong coupling expansion by $\beta$,
which is $0.037$ in the vacuum-suspended graphene.
(Here we take $\vf\sim c/300$ at any value of $\beta$.)
In this work, we only consider the leading order $[O(\beta^0)]$ in the strong coupling expansion,
which corresponds to the strong coupling limit ($\beta \rightarrow 0$) of the Coulomb interaction.
The gauge term $S_G$ does not contribute in this limit,
so that the partition function of this system can be written
only in terms of the fermionic term $S_F$ as
\begin{equation}
Z = \int [d a^\dag da] [db^\dag db] [dU_{\tau'}] e^{-S_F[a^\dag,a,b^\dag,b;U_{\tau'}]} .
\end{equation}

First we observe the behavior of the global symmetry of the system
when there is no lattice distortion.
By the integration over the link variables $U_{\tau'}$,
temporal kinetic terms [first two lines in Eq.(\ref{eq:lattice-action})] are converted into 4-Fermi terms
$-(1/4)n_a(x) n_a(x+\atau)$ and $-(1/4)n_b(x) n_b(x+\atau)$,
which are spatially local and temporally non-local by one lattice spacing.
Here the bosonic operator $n_\chi(x) \equiv \chi^\dag(x)\chi(x) \; (\chi=a,b)$
corresponds to the local charge density.
These terms can be considered as the on-site (Hubbard) interaction,
in which fermions on the same position with opposite spins interact with each other.
Such an interaction may lead to the spontaneous breaking of the sublattice symmetry
at the tree level,
as long as the spin symmetry is not broken.

Here we introduce an auxiliary field $\sigma \equiv \left\langle n_a - n_b \right\rangle$
by the Stratonovich--Hubbard transformation,
which corresponds to the charge density imbalance between two sublattices.
It corresponds to the ``chiral condensate'' $\langle \bar{\psi}\psi \rangle$,
the order parameter of chiral (sublattice) symmetry breaking, in the continuum effective theory.
By the mean-field approximation over $\sigma$,
the effective action at zero temperature is given in the quadratic form
\begin{equation}
S_F^{(0)} = \!\!\!\!\! \sum_{\bfr \in A \cup  B; \tau'} \!\!\!\!\! \frac{\sigma^2}{4} + \!\! \sum_{\bfk \in \Omega; \tau'} \!\!\! \Psi^\dag(\bfk,\tau') \!\!
  \left( \begin{array}{cc}
   \sigma/2 & \alpha \Phi^*(\bfk)\\
   \alpha \Phi(\bfk) & -\sigma/2
  \end{array}\right)
  \!\! \Psi(\bfk,\tau'), \label{eq:eff-action-1}
\end{equation}
where $\Psi \equiv (a,b)^T$ and $\alpha \equiv \atau h/\vf$.
Thus the integration over the fermions can be successfully performed,
yielding the free energy per one pair of A- and B-sites in the strong coupling limit,
\begin{equation}
\feff{0}(\sigma) = \frac{1}{2}\sigma^2 -\int_{\bfk \in \Omega} d^2\bfk \ln \left[\left(\frac{\sigma}{2}\right)^2 + \left|\alpha\Phi(\bfk)\right|^2\right]. \label{eq:feff_honeycomb}
\end{equation}
$\feff{0}(\sigma)$ is dominated by the second term (singularity from fermion one-loop) around $\sigma=0$,
while it is dominated by the first term (tree level of the auxiliary field) at large $|\sigma|$.
Therefore, the effective potential $\feff{0}(\sigma)$ has a minimum at finite $|\sigma|=0.343$,
which gives the charge density imbalance between two sublattices.
Finite $\sigma$ breaks the sublattice symmetry of the honeycomb lattice,
leading to a gap opening in the fermionic spectrum.
The term $(\sigma/2)(a^\dag a - b^\dag b)$ in Eq.(\ref{eq:eff-action-1}),
which corresponds to the effective mass term $m_{_\mathrm{eff}}\bar{\psi}\psi$
in the 4-component Dirac fermion description,
modifies the dispersion relation into $E(\bfk) = \pm\sqrt{|\Phi(\bfk)|^2 + (\vf \sigma/2\atau)^2}$.

\begin{figure}[tb]
\begin{center}
\includegraphics[width=7.5cm]{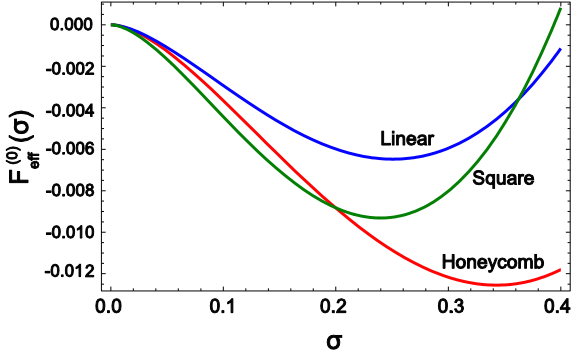}
\end{center}
\vspace{-0.5cm}
\caption{The free energy per one pair of A- and B-sites in the strong coupling limit, $\feff{0}$, as a function of $\sigma$.
``Honeycomb'' shows $\feff{0}(\sigma)$ in Eq.(\ref{eq:feff_honeycomb}) with the exact form of $\Phi(\bfk)$.
``Linear'' shows $\feff{0}(\sigma)$ in Eq.(\ref{eq:feff_honeycomb}) with the linear approximation $\Phi(\bfK_\pm + \bfk) \simeq (3/2)\ahc (\pm k_x+ik_y)$.
``Square'' shows the effective potential obtained by the strong coupling expansion with the square lattice formulation \cite{Araki_2010}
per four square lattice sites, corresponding to a pair of honeycomb lattice sites.}
\label{fig:feff_honeycomb1}
\end{figure}

In Fig.\ref{fig:feff_honeycomb1},
we compare the result from the honeycomb lattice formulation
with that from the square lattice formulation given in Ref.\onlinecite{Araki_2010}.
Here, $\feff{0}$ from the square lattice formulation is given by the free energy per four square lattice sites,
which corresponds to a pair of A- and B-sites on the honeycomb lattice.
We also show $\feff{0}(\sigma)$ in Eq.(\ref{eq:feff_honeycomb}) with the linear approximation of $\Phi(\bfk)$
in Fig.\ref{fig:feff_honeycomb1}.
Since they have the Dirac cone structure around $\bfk = \bfK_\pm$ in common,
the logarithmic divergence around $\sigma=0$ appears independently of the lattice formulation,
leading to the spontaneous sublattice symmetry breaking.
The quantitative difference in $\feff{0}$ comes from
the difference of the dispersion relation in the quadratic and higher order terms in momentum,
which are neglected in the linear approximation but become dominant far from the Dirac points in the momentum space.

Let us now examine the effect of an external Kekul\'{e} distortion on the sublattice symmetry;
it is represented by the additional Hamiltonian
\begin{equation}
H_K = \sum_{\bfr \in A} \sum_{i=1}^{3} \delta h_i(\bfr) a^\dag(\bfr) b(\bfr+\bfs_i) +\mathrm{H.c.}, 
\end{equation}
\begin{equation}
\delta h_i(\bfr) = \frac{\Delta}{3}\left[ e^{i(\bfK_+ \cdot \bfs_i+\mathbf{G} \cdot \bfr)} + e^{i(\bfK_- \cdot \bfs_i-\mathbf{G} \cdot \bfr)} \right], \label{eq:k-pattern}
\end{equation}
where $\mathbf{G} \equiv \bfK_+ - \bfK_-$,
and $\Delta$ denotes the amplitude of the external Kekul\'{e} distortion \cite{Chamon_2000}.
In this paper,
we consider the case in which the distortion amplitude $\Delta$ is real and uniform.
Since the distortion pattern is periodic by three times the unit cell of the honeycomb lattice,
the Brillouin zone $\Omega$ is split into three hexagonal regions,
$\tilde{\Omega}$ and $\tilde{\Omega}_\pm$,
surrounding $\bfk=0$ and $\bfk=\bfK_\pm$ respectively (see Fig.\ref{fig:brillouinzone2}).
For free fermions,
the Kekul\'{e} distortion opens a finite gap in the dispersion relation of quasiparticles,
$E_\mathrm{free}(\bfK_\pm+\bfk;\Delta) = \sqrt{|\bfk|^2+|\Delta|^2} +O(k^4)$,
while it preserves the sublattice symmetry of the honeycomb lattice.
The magnitude of the gap is proportional to the distortion amplitude $\Delta$.
In the Dirac fermion description,
Kekul\'{e} distortion can be written as an external field $\Delta\bar{\psi}\gamma_3\psi$,
which is invariant under the U(1) chiral transformation generated by the chirality (sublattice inversion) $\gamma_5$.

\begin{figure}[tb]
\begin{center}
\includegraphics[width=6cm]{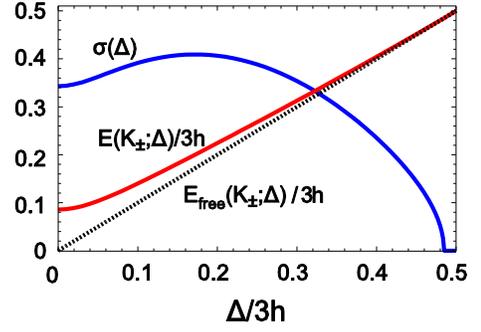}
\end{center}
\vspace{-0.5cm}
\caption{Charge density imbalance between two sublattices, $\sigma$,
and the total energy gap $E(\bfK_\pm;\Delta)=\sqrt{(\vf\sigma/2\atau)^2+\Delta^2}$,
as the functions of the external Kekul\'{e} distortion $\Delta$.
$\sigma(\Delta)$ first grows and then decreases as $|\Delta|$ increases,
vanishing at the critical value $\Delta_C/3h = 0.485$,
while $E(\bfK_\pm;\Delta)$ remains finite for any value of $\Delta$.
The dotted line shows the gap amplitude of the free fermion.}
\label{fig:sigma_delta}
\end{figure}

By adding the Kekul\'{e} term to the lattice effective action in Eq.(\ref{eq:eff-action-1}),
it is modified as
\begin{equation}
S_F^{(0)} \!\! = \!\!\!\!\! \sum_{\bfr \in A \cup  B; \tau'} \!\!\!\!\! \frac{\sigma^2}{4} + \!\! \sum_{\bfk \in \tilde{\Omega}; \tau'} \!\!\! \tilde{\Psi}^\dag(\bfk,\tau') \!\!
  \left( \begin{array}{cc}
   \!\! (\sigma/2) I_3 & \!\! \alpha \tilde{\Phi}^\dag(\bfk) \!\! \\
   \!\! \alpha \tilde{\Phi}(\bfk) & \!\! -(\sigma/2) I_3 \!\!
  \end{array}\right)
  \!\! \tilde{\Psi}(\bfk,\tau'), \label{eq:effaction-kekule}
\end{equation}
where $\tilde{\Psi}(\bfk) \equiv [ a(\bfk),a(\bfK_+ + \bfk),a(\bfK_- + \bfk),b(\bfk),b(\bfK_+ + \bfk),b(\bfK_- + \bfk) ]^T$
and $I_3$ is a $3\times 3$ unit matrix.
$\tilde{\Phi}$ is defined by a $3 \times 3$ matrix,
\begin{equation}
\tilde{\Phi}(\bfk) \equiv 
\left( \begin{array}{ccc}
   \Phi(\bfk) & \tilde{\Delta}\Phi(\bfK_- + \bfk) & \tilde{\Delta}\Phi(\bfK_+ + \bfk) \\
   \tilde{\Delta}\Phi(\bfK_- + \bfk) & \Phi(\bfK_+ +\bfk) & \tilde{\Delta}\Phi(\bfk) \\
   \tilde{\Delta}\Phi(\bfK_+ + \bfk) & \tilde{\Delta}\Phi(\bfk) & \Phi(\bfK_- +\bfk)
  \end{array}\right),
\end{equation}
with the dimensionless distortion amplitude $\tilde{\Delta} \equiv \Delta/3h$.
By integrating out the fermionic degrees of freedom,
we obtain the effective potential $\feff{0}$ as a function of $\sigma$,
with the external parameter $\tilde{\Delta}$.
When there is no Kekul\'{e} distortion,
it reproduces the effective potential in Eq.(\ref{eq:feff_honeycomb}).
{The minimum of the effective potential $\feff{0}(\sigma;\Delta)$
yields the expectation value of $\sigma$ as a function of $\Delta$,
as shown in Fig.\ref{fig:sigma_delta}.}
{By diagonalizing the matrix in Eq.(\ref{eq:effaction-kekule}) at $\bfk=0$,
we obtain the magnitude of the spectral gap as
\begin{equation}
E(\bfK_\pm;\Delta)=\sqrt{\left[\vf\sigma(\Delta)/2\atau\right]^2+\Delta^2},
\end{equation}
which is modified from the gap of the free electrons
by the distortion-dependent exciton condensate $\sigma(\Delta)$.}

As seen from Fig.\ref{fig:sigma_delta},
$\sigma$ grows around $|\Delta|=0$ and decreases at large $|\Delta|$,
with the critical value $|\Delta_C/3h| = 0.485$.
The reduction of $\sigma(\Delta)$ can be qualitatively understood
in terms of the effective field theory with a 4-component Dirac fermion:
the amplitude of the fermion propagator $(\bfk\cdot\mathbf{\gamma} + \sigma/2 + \Delta\gamma_3)^{-1}$
gets suppressed as $|\Delta| \rightarrow \infty$,
leading to a reduction of the dominance of the fermion one-loop effect in the effective potential.
The sublattice symmetry is fully restored, i.e. $\sigma$ vanishes,
at the critical value $\Delta_C$,
where $\sigma(\Delta)$ exhibits the mean-field critical exponent
$z=1/2$ with the definition $\sigma \propto |\Delta-\Delta_C|^z$.
As a result, we find that the external Kekul\'{e} distortion can
restore the sublattice symmetry of the honeycomb lattice,
although the gap in the spectrum remains finite,
{as shown in Fig.\ref{fig:sigma_delta}}.
On the other hand,
$\sigma(\Delta)$ increases quadratically as $\sigma(\Delta) = 0.343 + 7.02|\tilde{\Delta}|^2+O(\Delta^4)$
for small $|\Delta|$.
It comes from the quadratic and higher order terms in momentum
which are neglected in the continuum effective theory;
if the linear approximation is applied to $\feff{0}(\sigma;\Delta)$,
the resulting $\sigma(\Delta)$ decreases monotonically around $\Delta=0$.

In this paper,
we have investigated the interplay between two symmetry breaking patterns
of the honeycomb lattice,
in the strong coupling limit of the Coulomb interaction.
Due to the on-site part of the Coulomb interaction,
the sublattice symmetry is spontaneously broken
if the system has the perfect honeycomb lattice symmetry.
As we introduce the Kekul\'e distortion as an external field,
the system reveals a second-order phase transition --
restoration of the sublattice (chiral) symmetry.
As a result,
the gap amplitude of monolayer graphene in the strong coupling limit is modified
from that of free fermions
if the external Kekul\'e distortion is under the critical value,
while it is not modified at large Kekul\'e distortion.
If the ``chiral symmetry broken'' phase is achieved experimentally in the vacuum-suspended graphene,
the modification of the gap amplitude will have an effect
on gap engineering,
which has recently become important for industrial applications
as electronic devices.



While the Kekul\'{e} distortion is introduced as an external parameter in the present paper,
it is proposed to occur spontaneously
if the interaction between the nearest-neighboring (NN) or the second-NN sites is taken into account
\cite{Nomura_2009,Raghu_2008,Weeks_2010}.
We are currently studying such an effect
in the next-to leading order $[O(\beta^1)]$ in the strong coupling expansion,
which includes the NN interaction
$ \sum_{j=1}^{3}\left[a^\dag(x)b(x+\bfs_j)b^\dag(x+\bfs_j+\atau)a(x+\atau) +\mathrm{H.c.}\right] $.
With a sufficiently strong NN interaction,
our preliminary study shows a phase transition between the sublattice symmetry-broken phase
and the spontaneous Kekul\'{e} distortion phase.
Such a phase transition is expected to be of first order,
since either the chiral condensate or the Kekul\'e distortion remains finite
due to the logarithmic behavior in the effective potential.
The effect of the phonon-mediated interaction is also of great importance;
if such an interaction is taken into account,
the superconducting order may also have to be considered
\cite{Khveshchenko_2009,Nunes_2010}.

\
\begin{acknowledgments}
The author thanks H.~Aoki, C.~DeTar, T.~Hatsuda, K.~Nomura and S.~Sasaki for valuable comments and discussions.
This work is supported by Grant-in-Aid for Japan Society for the Promotion of Science (DC1, No.22.8037).

\end{acknowledgments}


\end{document}